\documentclass[letterpaper,twocolumn,fleqn]{article} 

\usepackage{ist}
\usepackage{amsfonts}

\title{Per Clip Lagrangian Multiplier Optimisation for HEVC}

\author{ Daniel J Ringis, Fran\c{c}ois Piti\'e and Anil Kokaram; Sigmedia Group,  Electronic and Electrical Engineering Dept., Trinity College Dublin; Dublin, Ireland}

\date{} 

\begin{document} 

\maketitle 

\thispagestyle{empty} 


\begin{abstract}
The majority of internet traffic is video content. This drives the demand for video compression in order to deliver high quality video at low target bitrates.
This paper investigates the impact of adjusting the rate distortion equation on compression performance.
A constant of proportionality, k, is used to modify the Lagrange multiplier used in H.265 (HEVC). Direct optimisation methods are deployed to maximise BD-Rate improvement for a particular clip.
This leads to up to 21\% BD-Rate improvement for an individual clip. Furthermore we use a more realistic corpus of material provided by YouTube. The results show that direct optimisation using BD-rate as the objective function can lead to further gains in bitrate savings that are not available with previous approaches.
\end{abstract}

\section{Introduction}
\label{sec:intro}
Video content is predicted to rise to over 80\% of internet traffic by the year 2021 \cite{cisco}. This continues to drive video compression technology. A successful video codec would provide a bitstream which is capable of being decoded at the highest possible quality for a target bitrate, or the lowest possible bitrate for a target quality or distortion. One of the key challenges to be solved in the design of a practical codec is the tradeoff between rate and distortion. A Lagrange multiplier approach was advocated by Sullivan et al \cite{sullivan1998rate} and that has been the adopted view since 1998. In this approach the codec makes a number of decisions in order to minimise a cost \textit{J} as follows. 
\begin{equation}
    J = D + \lambda R \label{rd}
\end{equation}
As can be seen $J$ combines both a distortion \textit{D} (for a frame or macroblock) and a rate \textit{R} (the number of coded bits for that unit) through the action of the Lagrangian multiplier $\lambda$.
Different choices for $\lambda$ result in different R/D tradeoffs. The idea is used across a wide range of codec operations e.g. block type (Skipped/Intra/Inter), motion vectors and bit allocation at the frame and clip levels \cite{wiegand2001lagrange}.

The choice of \(\lambda\) was determined experimentally in the early days of codec research \cite{ortega1998rate}. This parameter can be considered as a kind of hyperparameter in a hybrid codec as it affects so many other decisions. This was originally implemented for the H.263 codec and then modified for H.264 and H.265 (HEVC) \cite{hevcOverview}. The approach was to select a value based on a cohort of examples such that the performance was optimised {\em on average} across the set. That set was quite small, less than 5 examples. More recently, Katsavounidis et al \cite{netflix}\cite{katsavounidis2018video} recognised that by choosing codec hyperparameters {\em per clip} significant gains could be achieved. This is because the statistics of video clips varies greatly over any corpus. While that work considered only the design of a bitrate ladder for DASH streaming, it provided an impetus for {\em per clip} optimisation throughout the codec system. Covell et al\cite{covell2016optimizing} also built on that idea by optimising other codec hyperparameters (CRF) in that case for {\em per clip} transcoding.

\begin{figure}
    \centering
    \includegraphics[width=\columnwidth]{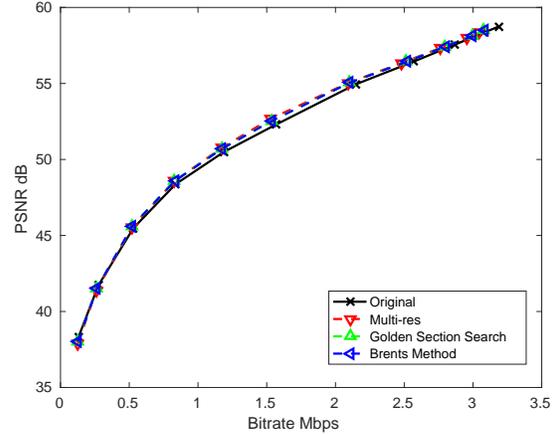}
    \caption{Rate Distortion curves for sequence:  Gaming\_360P-1e31. The RD curves generated using any one of the proposed algorithms outperforms the curve for X265 using the default $\lambda$ (black).}
    \label{idealRDcurves}
\end{figure}

This paper explores the idea that a better \(\lambda\) exists for an individual video clip. Unlike previous work, \(\lambda\) was adjusted using direct optimisation techniques using BD-Rate as the objective function. In addition, a more comprehensive and modern corpus of test data was used, provided recently by YouTube \cite{wang2019youtube}. For a given clip it was found that up to a 21\% improvement can be had by adjusting the Lagrangian multiplier in the rate distortion equation. An example of the gains reported in this paper are shown in Figure~\ref{idealRDcurves}. The proposed algorithms all result in RD performance that is well above the default behaviour in the X265 codec (in this case). This result is discussed in more detail later, but it contrasts well with previous work that reports up to 11\% improvement. The next section presents more detailed background and a review of recent work.

\section{Lagrangian Multiplier in Compression}

There has been a tremendous amount of work done to establish what the Lagrangian multiplier should be in the different iterations of the MPEG codecs (H.263, H.264 and H.265). Being able to find the balance between the quality of decoded images and channel capacity is the fundamental problem which the rate distortion algorithm attempts to solve \cite{wiegand1996rate}. 

As alluded to in the introduction, the seminal work of Sullivan and Wiegand\cite{sullivan1998rate} laid the foundation for an experimental approach to choosing an appropriate $\lambda$. In that work, they established the relationship between quantisation step size $Q$ (used for DCT coefficient quantisation) and the distortion $D$ in a frame i.e. $D(Q) = a Q^2$, where $a$ is a constant relating to the content.  Minimising $J$ wrt $D$ then yields a relationship $\lambda = D/b$ where $b$ is another content related constant. Combining these two relationships then leads to a relationship between $\lambda$ and $Q$ as follows
\begin{equation}
   \lambda = 0.85 \times Q^2 \label{lambdaQ}
\end{equation}
where the constant was determined experimentally over a small corpus of 3 video clips (CIF resolution). This was explicitly presented for H.263 in 2001 \cite{wiegand2001lagrange}.

Updates to these experiments then yielded similar relationships for H.264 and H.265 (HEVC).  Because of the introduction of bi-directional (B) frames the constants are all different. In fact three different relationships were established for each of the Intra (I), Predicted (P) and B frames as follows.
\begin{equation}
   \lambda_{I}=0.57 \times 2^{(Q-12)/3} \label{lambdaI}
   \end{equation}
   \begin{equation}
\lambda_{P} = 0.85 \times 2^{(Q-12)/3} \label{lambdaP}
\end{equation}
\begin{equation}
\lambda_{B}= 0.68 \times \max(2, \min(4, (Q - 12)/6)) \times 2^{(Q-12)/3} \label{lambdaB}
\end{equation}

Clearly the size of the corpus used for these experiments was quite small, so there is room for exploration with a larger corpus. In addition, average performance is known to be worse than individual adaptation as discussed previously. Hence the following section outlines different methods which have been used to determine a better Lagrangian multiplier for use in video encoding.

\subsection{Adaptive Lagrange Multipliers}

There have been a few attempts to adjust the rate distortion optimisation within a video codec. In all cases, the authors have discovered that a better $\lambda$ does exist {\em per clip}.
The techniques are best considered as two classes. In one class (M-Algorithms), $\lambda$ is adjusted away from the default stated previously, using a constant of proportionality $k$ as follows
\begin{equation}
   \lambda_{\textrm{new}} = k \times \lambda_{\textrm{orig}}\label{kfactor}
\end{equation}
where $\lambda_{\textrm{orig}}$ is the default Lagrangian multiplier estimated in the video codec, and  $\lambda_{\textrm{new}}$ is the updated Lagrangian.  Algorithms in this class, learn a relationship between $k$ and features extracted from the video stream. 
Zhang and Bull  \cite{zhang_bull} used a single feature ${D}_{P}/{D}_{B}$, the ratio between the MSE of P and B frames. This feature gives some notion of temporal complexity. Experiments based on the DynTex database yielded a choice for $k$ as follows.
\begin{equation}
    k = a({D}_{P}/{D}_{B}+d)^b +c\label{zhangbull}
\end{equation}
where \(a, b, c \textrm{ and } d \) were experimentally calculated and different for each codec tested (H.264 and H.265). In their work, $k$ was updated every GOP and they report up to 11\% improvement in BD-Rate. This approach of fitting a simple curve was also used by Yang et al, except that they used a different feature, a perceptual content measurement $PC$ as follows
\begin{equation}
   k = a\textrm{PC} -b \label{yang}
\end{equation}
where again $a,b$ were determined experimentally using a corpus of the Derfs dataset. Unlike Zhang and Bull, this was a straight line fit, the loss in complexity of the fit is compensated for by the increase in complexity of the feature. 

Ma et al \cite{ma2016adaptive}  used four features with a SVM to determine $k$. The SVM was determined using the DynTex dataset. They reported up to 2dB improvement in PSNR and 0.05 improvement in SSIM at equal bitrates.  The four video features measured spatial and temporal complexity directly from texture and motion measurements on the pixel level data itself. Hamza et al \cite{hamza2019parameter} also take a Machine learning based approach but here using a semantic scene model. Their work used the Derfs dataset and reported up to 6\% BD-Rate improvement.

Since $\lambda$ is linked to $Q$ as explained previously, the second class of algorithms (Q-algorithms) adjusted $\lambda$ implicitly by adjusting the quantiser parameter $Q$. Papadopoulos et al\cite{Papadopoulos} applied an offset to Q, in HEVC, based on the ratio of the distortion in the P and B frames. Each QP was updated from the previous Group of Pictures (GOP) using the following equation:
\begin{equation}
    \textrm{QP} = a \times \textrm{MSE}_{\textrm{ratio}} - b \label{qpupdate}
\end{equation}
where $a,b$ are constants determined experimentally.
 This lead to an average BD-Rate improvement of 1.07\% on the DynTex dataset. With up to 3\% BD-Rate improvement achieved for a single sequence.
Taking a local, exhaustive approach, Im and Chan \cite{ImChan} proposed encoding a frame multiple times in a video. Each frame was encoded using a Quantiser Parameter from the set (\( QP \in ({QP, QP \pm 1, QP \pm 2, QP \pm 3, QP \pm 4}) \)). This led to an increased complexity in coding in HEVC, however for a single sequence achieved a 14.15\% BD-Rate improvement.

However, both M/Q-Algorithms have overlooked the possibility that a direct search using BD-Rate as the objective function, could lead to even more improvements. In some way all these previous works have reported data showing $k$ versus $BD-Rate$, but they use various summary models to simplify this relationship across a corpus. By using instead Direct optimisation, we are able to explore just how much more BD-Rate gains there are per-clip. This idea is presented next.

\begin{figure}
    \centering
    \includegraphics[width=0.85\columnwidth]{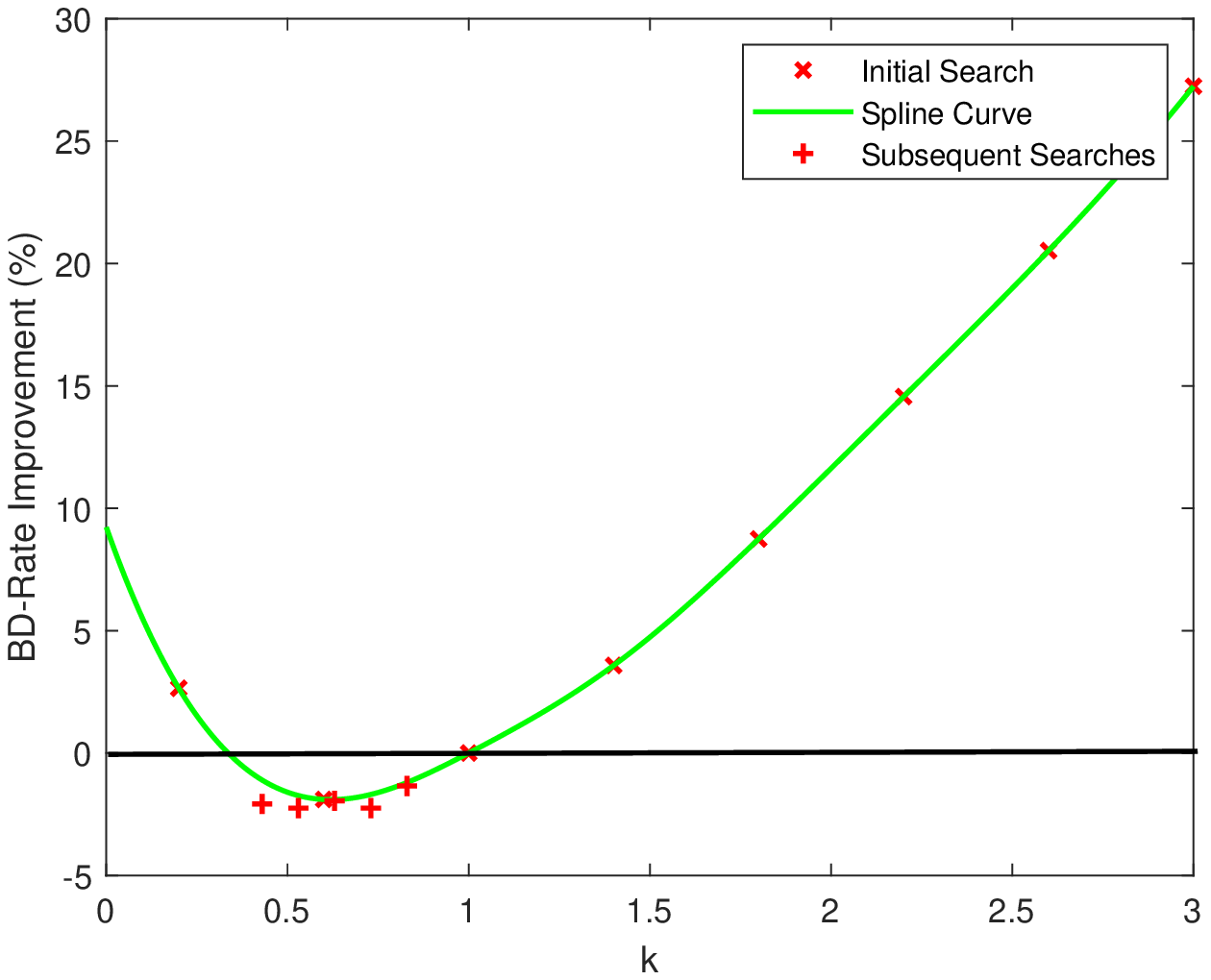} 
    \includegraphics[width=0.85\columnwidth]{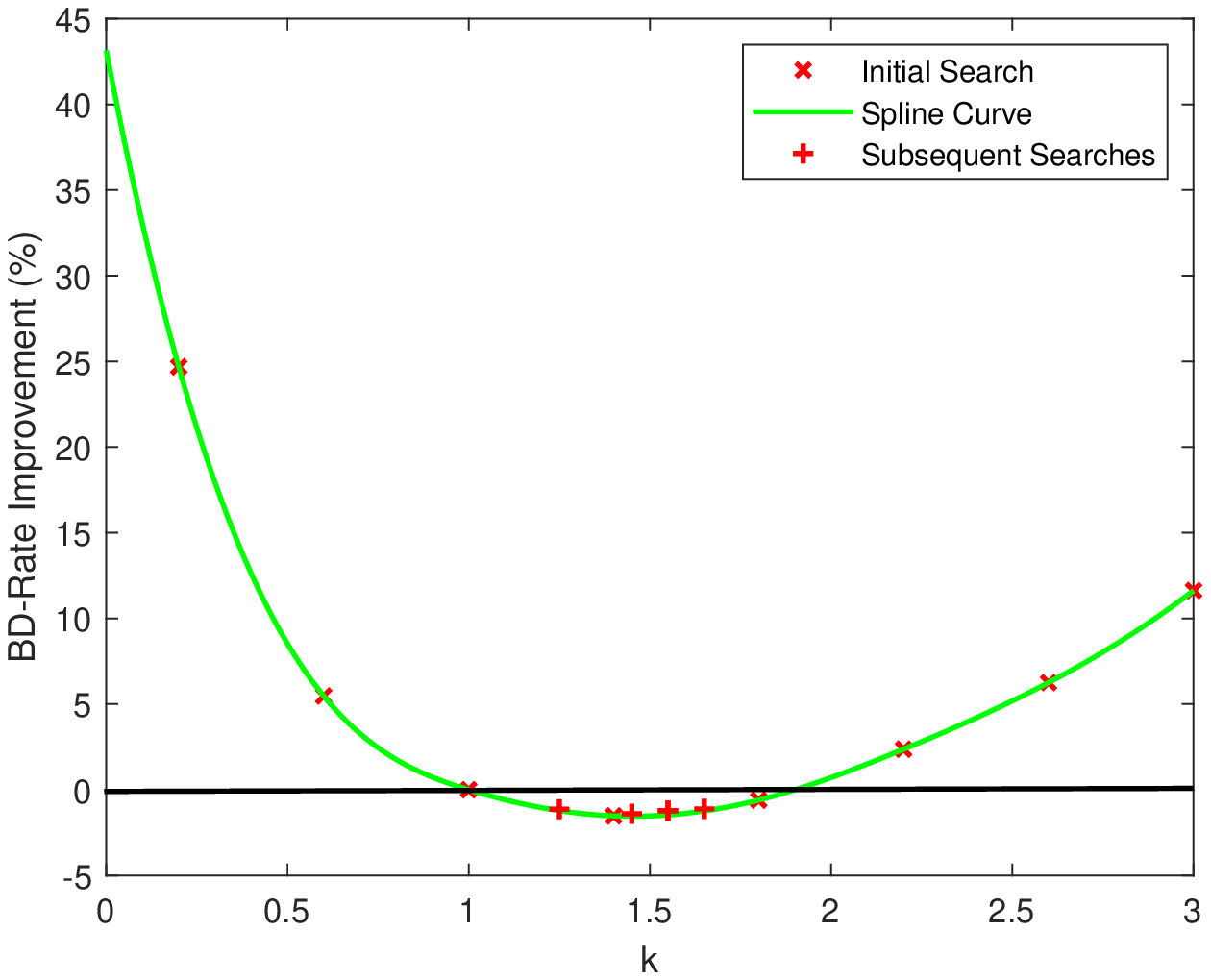}
      \includegraphics[width=0.85\columnwidth]{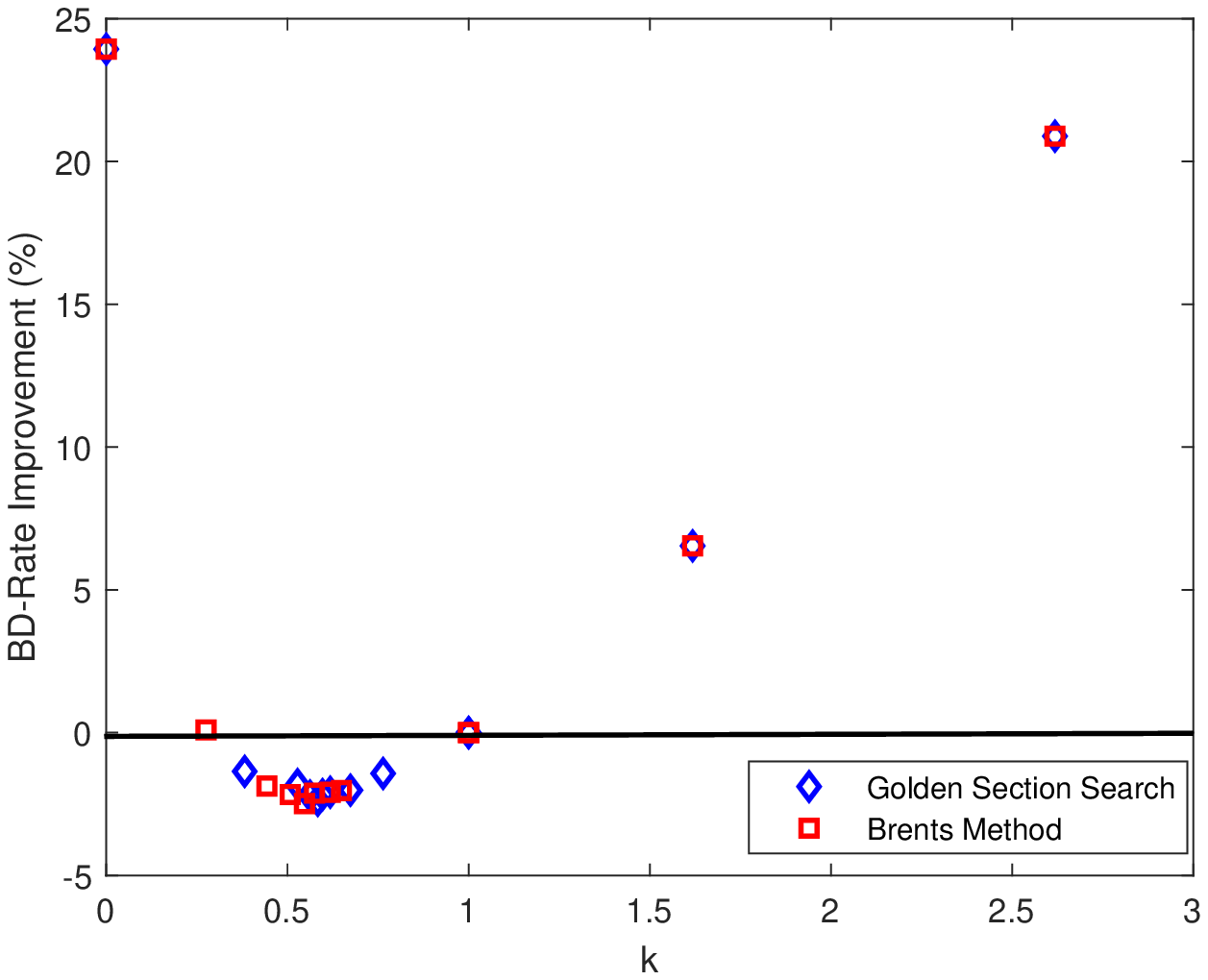} 
    \includegraphics[width=0.85\columnwidth]{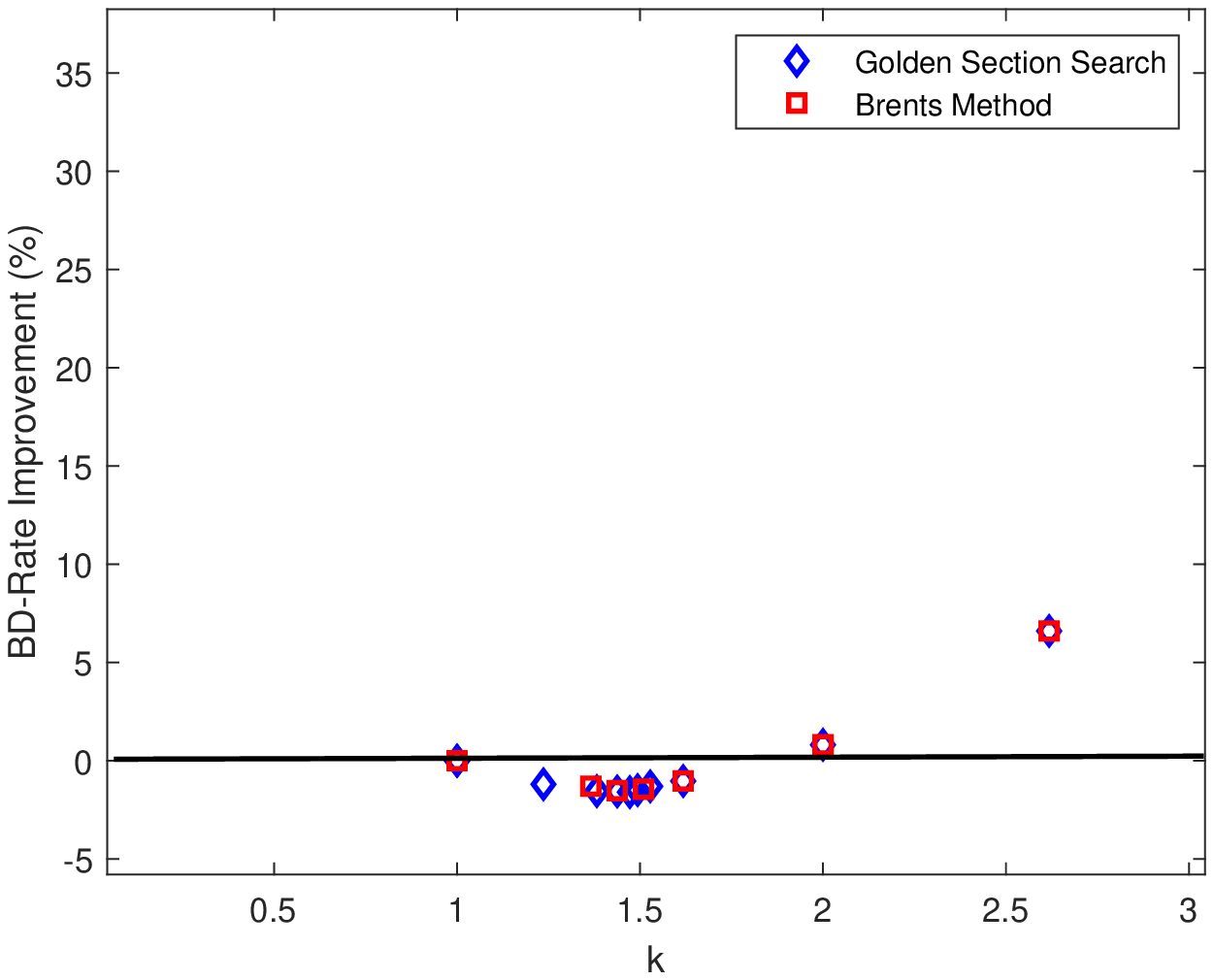}
    \caption{BD Rate(\%) vs $k$ using two clips ((NewsClip\_720P-7745 and CoverSong\_720P-3dca). Top : Multi-Res search, where X's are the initial search results, the solid line is the fit spline interpolation and +'s are the subsequent searches. Bottom : Brents Method $\Diamond$ and Golden Section $\Box$.}
    \label{examplecurve}
\end{figure}{}

\section{Direct Optimisation}
The approach adopted here falls under the M-algorithm class discussed above. Hence an estimate of $k$ is required such that $ \lambda_{\textrm{new}} = k \times \lambda_{\textrm{orig}}$. In this work however, we directly minimise BD-Rate with respect to $k$ using a range of optimisation strategies. Furthermore, we deploy the optimisation strategies using a modern dataset of User Generated content containing realistic material provided by YouTube \cite{wang2019youtube}. The new objective function is now directly the BD-Rate \cite{bdrate} as a function of $k$, $B_r(k)$ as follows.
\begin{equation}\label{bjontegaardEqn}
    B_r(k) = \int_{D_1}^{D_2} (R_1(D) - R_k(D)) dD
\end{equation}
where the integral is evaluated over the quality range $D_1..D_2$. $R_1(D), R_k(D)$ are the RD curves corresponding to the default ($k=1$) and the evaluated multiplier $k$ respectively. Each RD operating point is generated at a constant bitrate within a range which matches typical streaming media use cases.  The overall optimisation process is then as follows.
\begin{enumerate}
    \item Generate an RD curve using $\lambda_{orig}$ and $R_t = 256k : 7M$.  We use 11 operating points in this range.
    \item Define our BD-Rate objective function as specified above in equation \ref{bjontegaardEqn}. We use the same polynomial-log fit for evaluating the integral as recommended in \cite{bdrate}. 
    \item Starting from $k=1.0$,  minimise  $B_r(k)$, BD-Rate, wrt $k$ using any typical optimisation routine.
\end{enumerate}
Note that for every evaluation of $B_r(k)$ , eleven (11) encodes are required as well as the subsequent BD-Rate calculation itself.

\subsection{Optimisation Algorithms}
Three simple direct optimisation algorithms were tested. The first used a multi-resolution search (multi-res). This is a simple brute force strategy in which successively refined grid positions in $k$ were searched. The initial grid was \(k = 0.2:0.4:3.0\). The value of $k$ which yielded the minimum $B_r(k)$, $k_{\textrm{opt}}^1$ was calculated based on a spline curve fit to the sparse points, evaluated at steps of $0.01$. Successive refinements searching positions of $k_{\textrm{opt}}^n\pm \Delta$ were then performed with $\Delta = 0.2, 0.1, 0.05$ in succession.
Figure ~\ref{examplecurve} shows this process in action using two clips. X's indicate the initial evaluated positions, and the solid line is the initial curve fit. The subsequent refinement grids are indicated with +. This process therefore always required 15 evaluations of the objective function.

 Golden Section Search and Brent's method \cite{NumericalMethods} were used as the other two optimisation schemes. These are smarter direct search methods using local curve fits. As the multi-res method required fifteen (15) evaluations of the objective function, these optimisation routines were terminated after fifteen iterations and a tolerance of 0.02\% in the objective function. This allowed a fair comparison. 
 
 An example of the search results ($k$ vs BD-Rate) for the methods can be seen in Figure \ref{examplecurve} for two particular clips. Implementation details are in the next section, but one key observation here is the very different appearance of each curve. This shows that a summary model (e.g. used by Zhang and Bull) may be unable to exploit all the gains available. Furthermore, we note that the points discovered by the direct search methods deviate from the initial smooth curve fit near the minimum. This shows the complexity of the surface in the region of interest. Finally, note that the initial iterations of Brent's and Golden searches cover a very similar curve as the Multi-res approach. This means that the complexity of the curves is actually limited to the region near the minimum.

\subsection{Implementation details}
For this work VideoLAN's \cite{videoLAN} implementation of H.265 was used , {\tt x265}\footnote{Version: 3.0+28-gbc05b8a91}. The codebase was modified to take as an argument, the multiplier $k$ and hence alter the Lagrangian constraint \ref{kfactor} used in RD control throughout the codec.

In order to generate Rate Distortion curves (RD-curve), each video was encoded using the default input parameters for {\tt x265} at a constant bitrate. The target bitrate was set within the range of 256k to 7M. Both SSIM and PSNR were used as the distortion metric. The command invocation used for each video encode was as follows:

\texttt{x265 --input SEQ.y4m --bitrate <XXXX> --tune-<YYYY> --<YYYY> --csv-log-level 2 --csv DATA.csv --output OUT.mp4 }

Where SEQ, OUT and DATA are the filenames for the raw input file, output encoded video and the log file for an encode respectively. XXXX is the target bitrate and YYYY is the distortion metric used, either PSNR or SSIM.

\subsection{Dataset}
\begin{figure}
    \centering
    \includegraphics[width=0.28 \columnwidth]{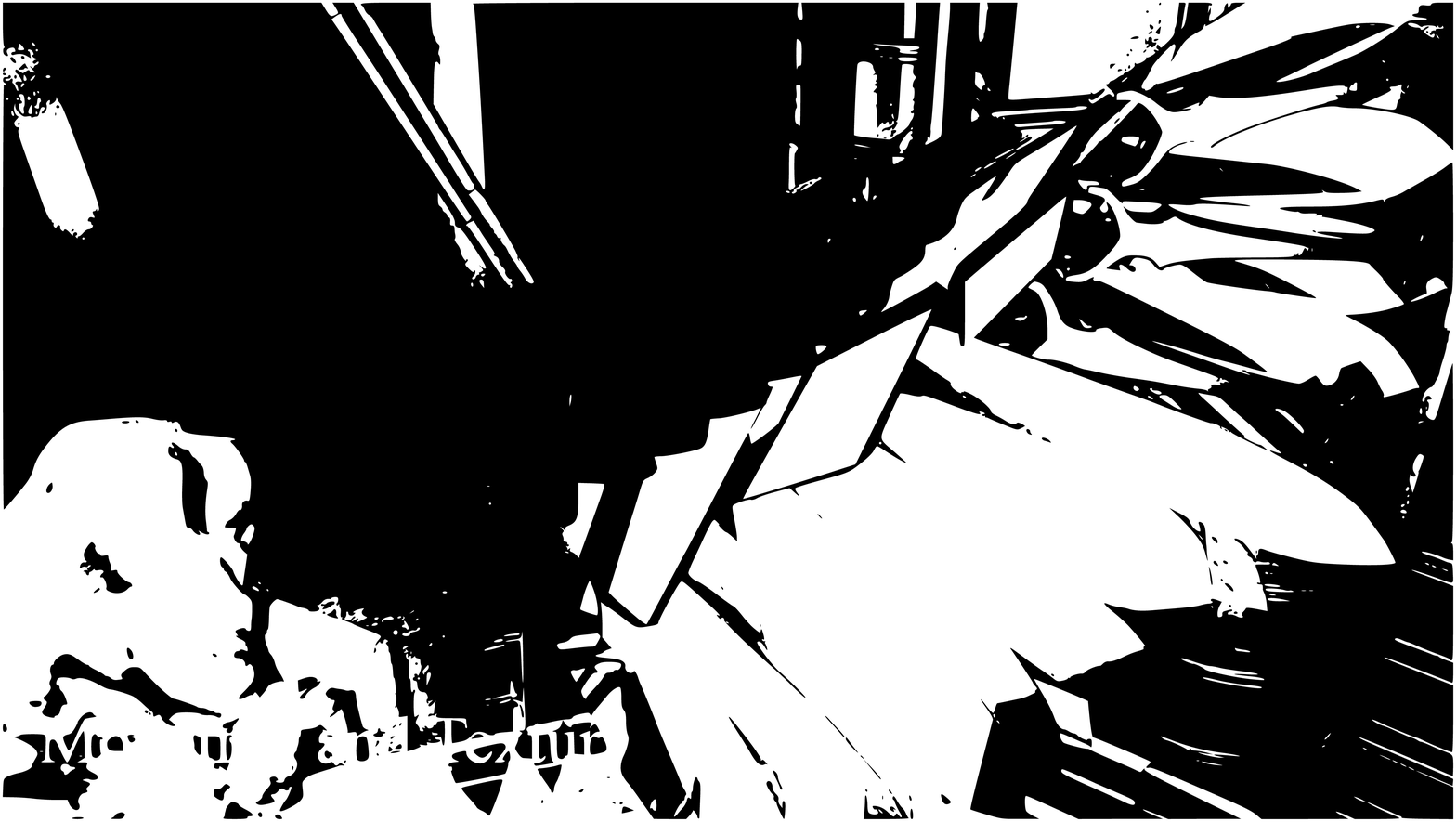}
    \includegraphics[width=0.28 \columnwidth]{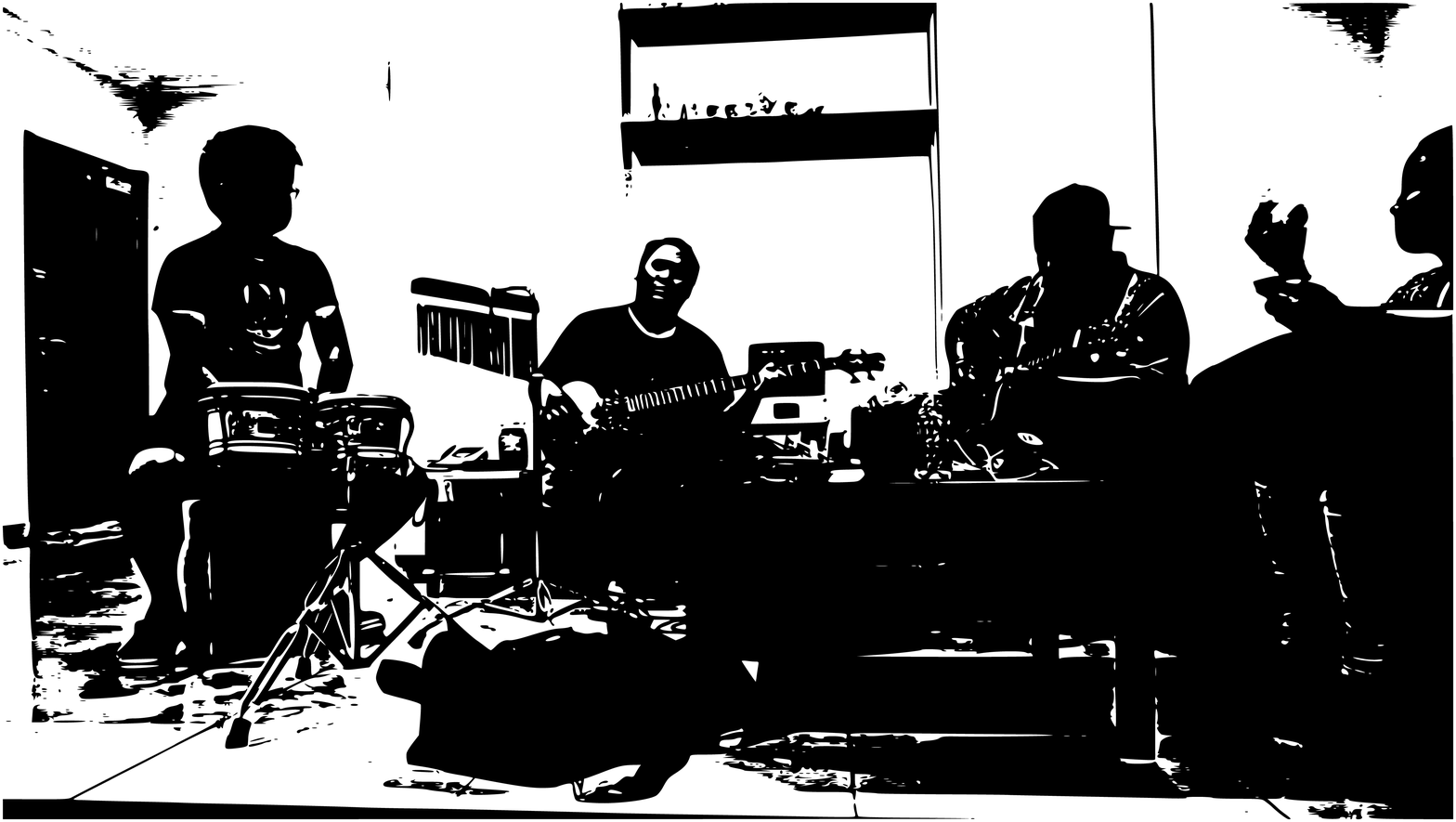}
    \includegraphics[width=0.28 \columnwidth]{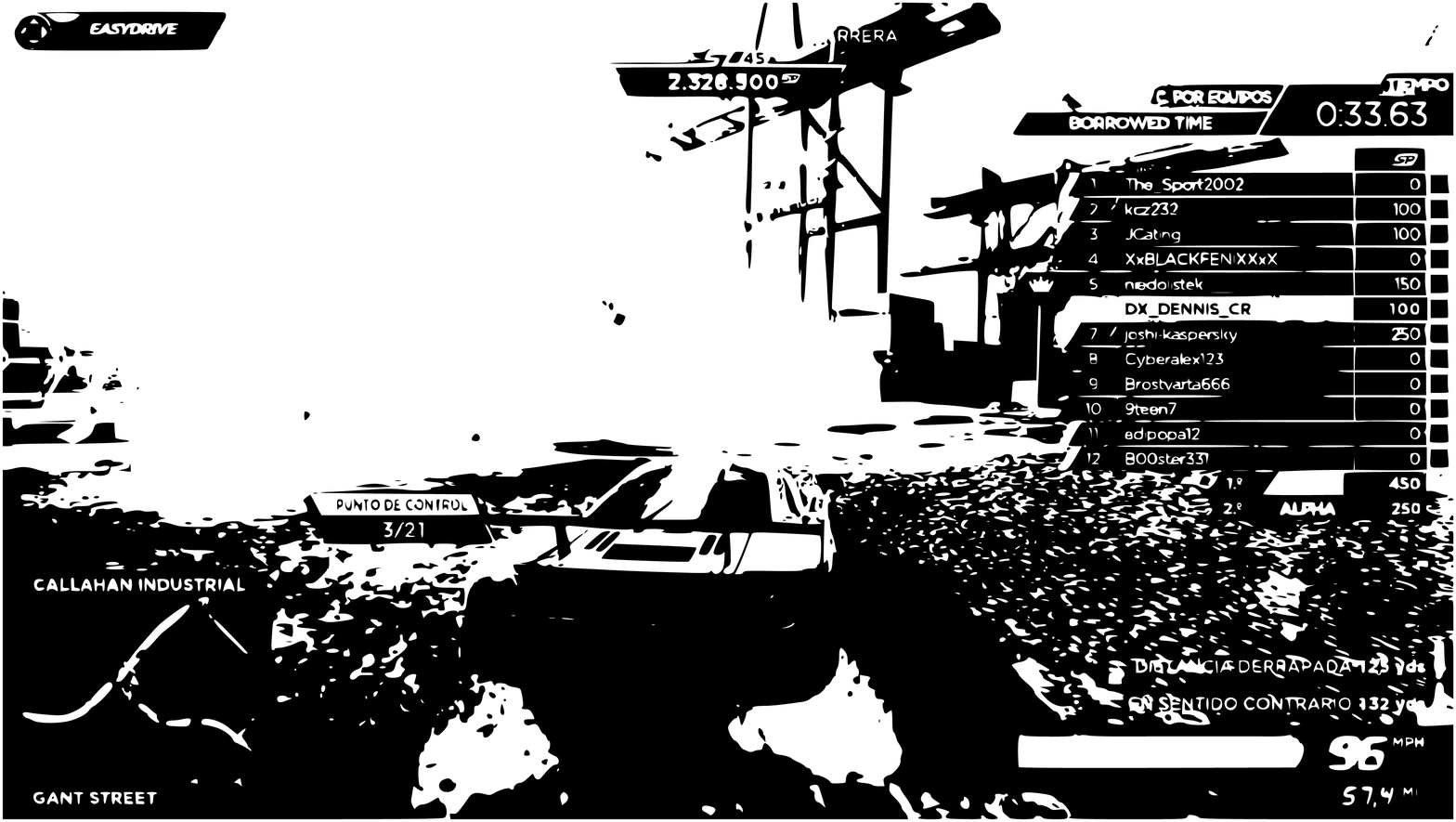}
    \\
    \includegraphics[width=0.28 \columnwidth]{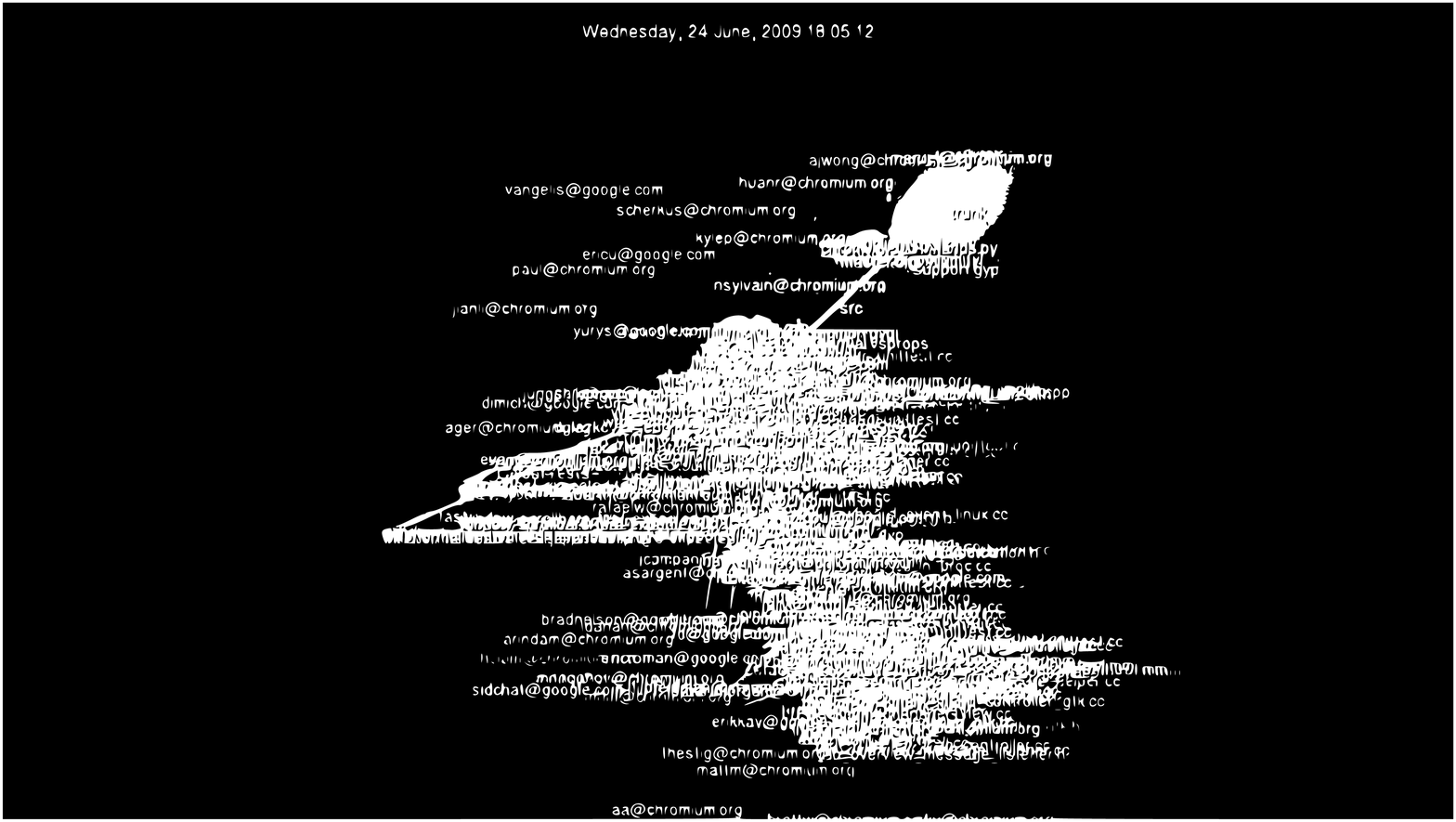}
    \includegraphics[width=0.28 \columnwidth]{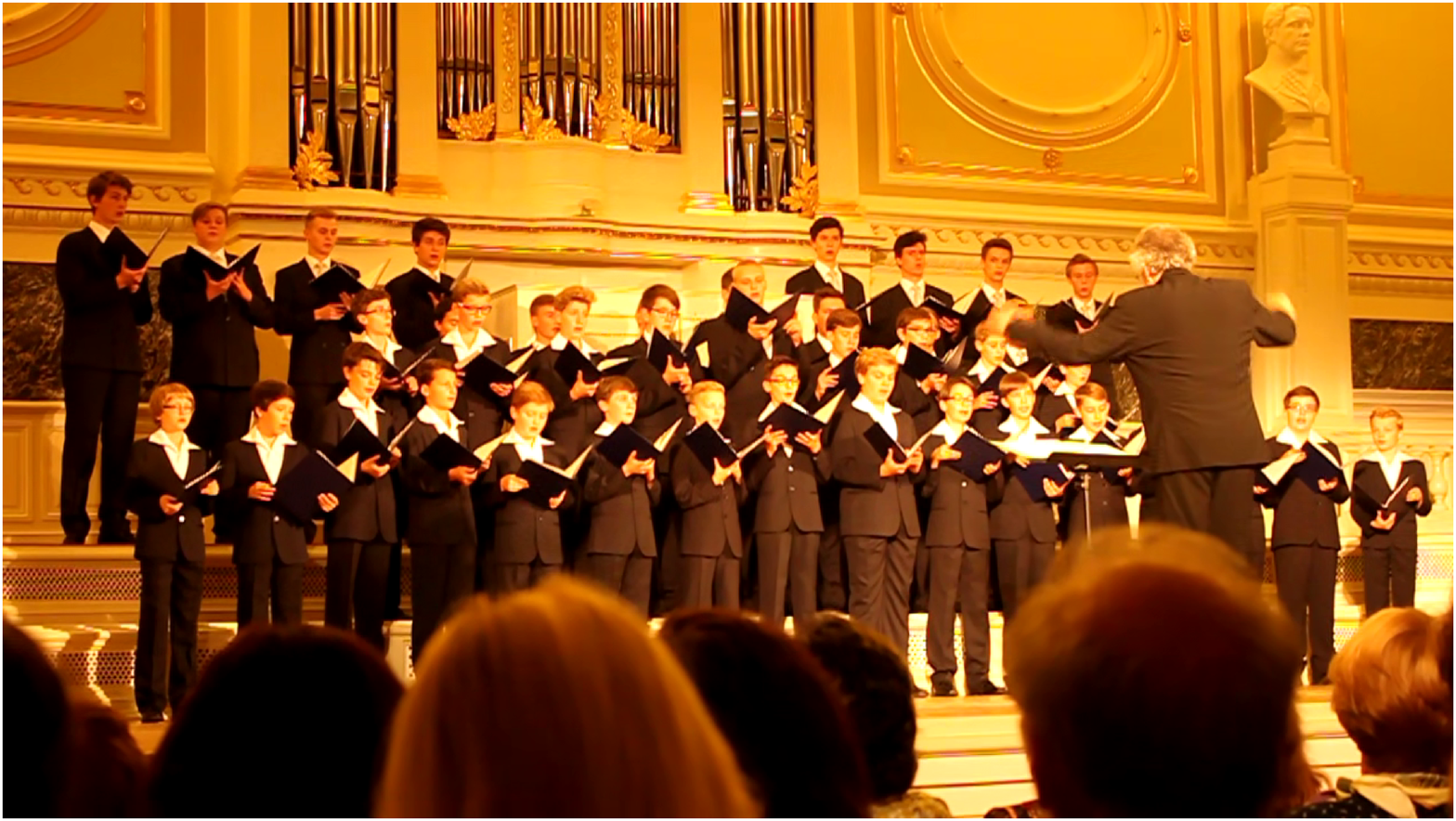}
    \includegraphics[width=0.28 \columnwidth]{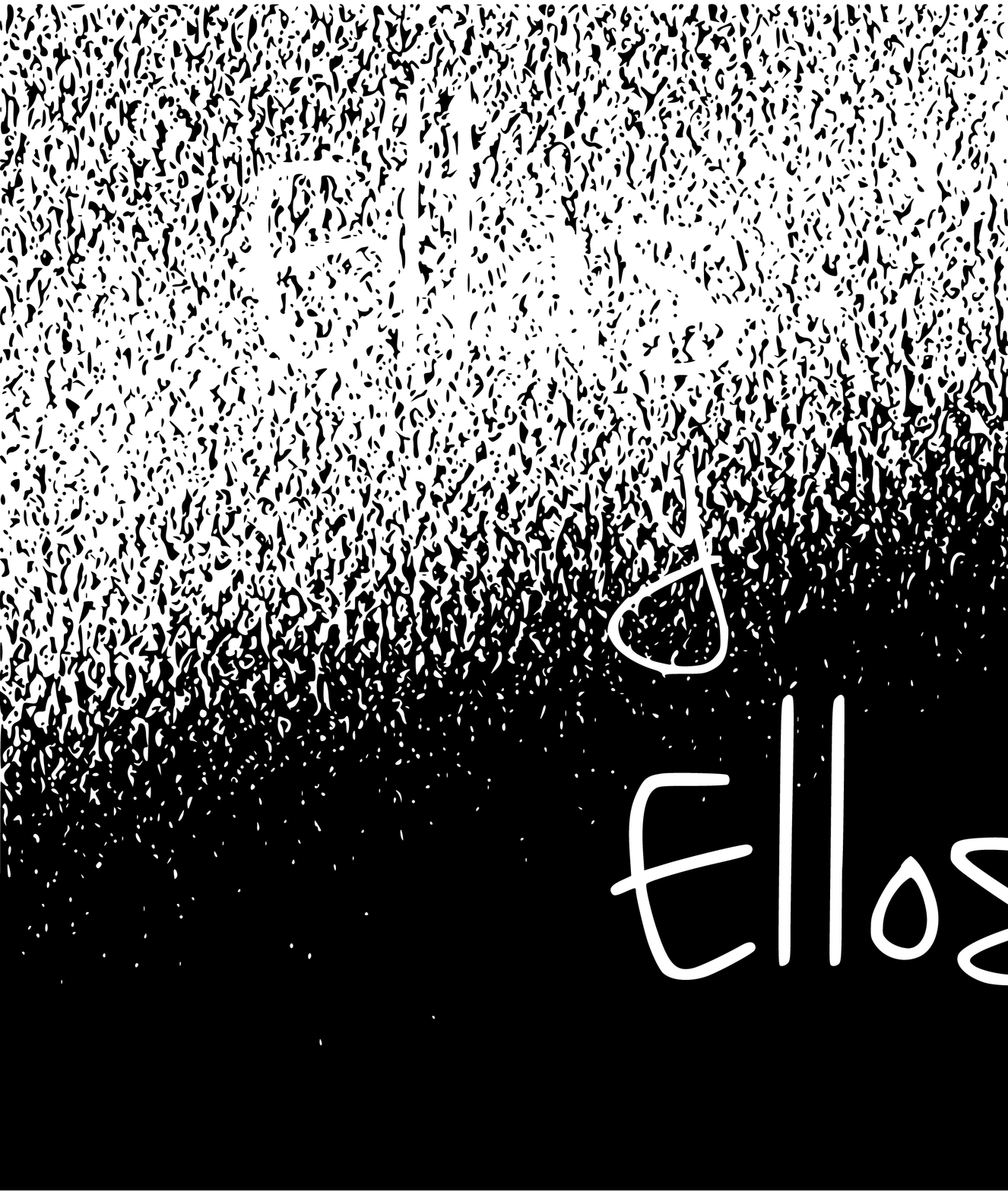}
    
    \caption{Example frames from each class in the YouTube dataset.}
    \label{datasetExample}
\end{figure}

Previous work in this space \cite{zhang_bull}\cite{ma2016adaptive}\cite{Papadopoulos} has used generally low resolution video with content that does not reflect current use.  A sub-sample of videos from the recently released YouTube-UGC Dataset\footnote{media.withyoutube.com} \cite{wang2019youtube} was therefore selected for use in this paper\footnote{The specific location of the chunks in the clips provided is available: http://www.mee.tcd.ie/~sigmedia/EI2020}. This should provide a better representation of widely used and viewed video content.
Each video clip used was 150 frames (5 seconds) in length to match the chunks typically used by Dynamic Adaptive Streaming over HTTP (DASH) systems. The selected clips contained Forty-four (44) at 720p resolution and thirty-three (33) clips at 360p resolution, all at 30 fps. These clips represented eleven (11) classes of video as specified by the YouTube team. A sample of each class of the clips can be seen in Figure \ref{datasetExample}.

\section{Results}
\begin{figure*}
    \centering
    \includegraphics[width=2\columnwidth]{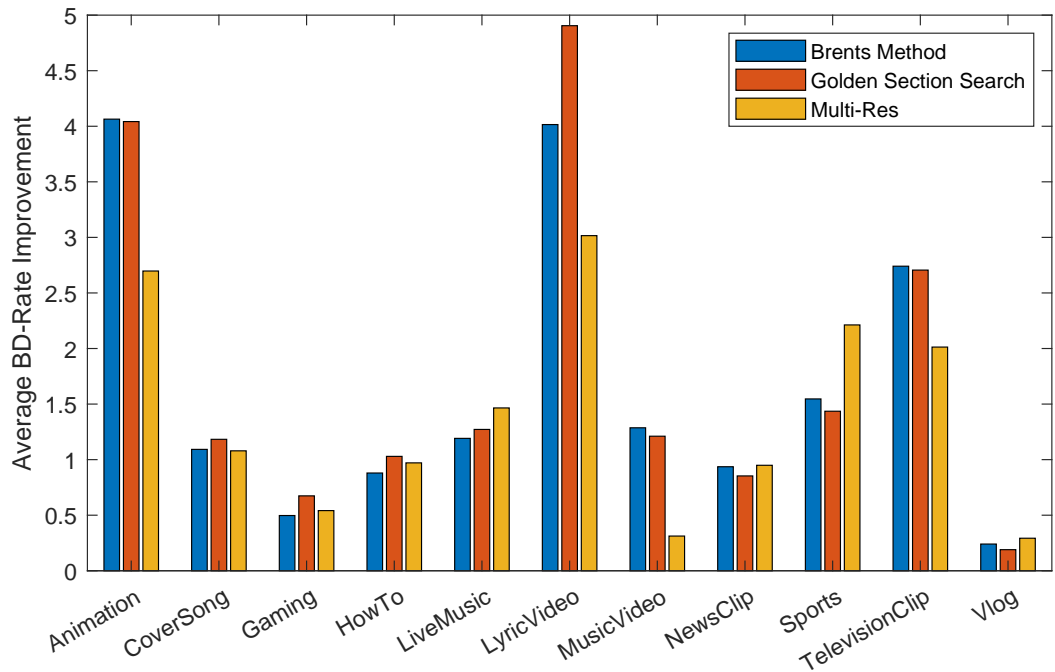}
    \caption{Average BD-Rate improvement (\%) for each class of video. Both Brent's Method and Golden Section Search require less iterations than Multi-Res search and often provide better BD-Rate improvement.}
    \label{avgResults}
\end{figure*}

Figure \ref{avgResults} shows the average BD-Rate improvement for each sequence class for each of the three direct optimisation methods. The bar chart shows that in almost all cases, Brent's and Golden methods are better than multi-res. Per class gains in BD-Rate are within 0.5\%-5\%. For three classes (Live, Sports and Vlog) multi-res is best. This is probably due to the restrictions on maximum allowed iterations in the competing methods. The best BD-Rate improvement for each class is listed in Table \ref{allResults}. Up to 21\% is observered in an animation clip. The corresponding BD-Distortion is less than 0.01dB. This highlights that the gains are all in bitrate without loss in quality, and that there are significant gains to be had {\em per clip}.

\begin{table}[!h]
\caption{Table \ref{allResults}: Best BD-Rate (\%) improvement for each class of sequence using the three direct optimisation methods}
\label{allResults}
\begin{center}       
\begin{tabular}{|l|l|l|l|}
\hline \textbf{Class}          & \textbf{Brent} & \textbf{Golden}& \textbf{Multi-Res}\\ \hline
Animation      & \textbf{21.667} & \textbf{21.667}    & 13.109            \\ \hline
CoverSong      & 1.839             & 2.279              & \textbf{2.377}    \\ \hline
Gaming         & \textbf{1.858}  & 1.762              & 1.665             \\ \hline
HowTo          & 5.023           & \textbf{5.239}      & 5.037            \\ \hline
LiveMusic      & 1.570           & \textbf{1.810}     & 1.776             \\ \hline
LyricVideo     & \textbf{15.879} & \textbf{15.879}    & 10.383            \\ \hline
MusicVideo     & 6.417           & \textbf{6.760}     & 0.516             \\ \hline
NewsClip       & \textbf{2.481}  & 2.371              & 2.251             \\ \hline
Sports         & \textbf{8.391}  & 7.583              & 8.378             \\ \hline
TelevisionClip & \textbf{10.533} & 10.238             & 9.486             \\ \hline
Vlog           & 0.351           & 0.496              & \textbf{0.590}    \\ \hline                                                
\end{tabular}
\end{center}
\end{table} 

On average golden section and Brent's search took 12.6 and 9.7 iterations respectively. This was with a 0.02\% BD-Rate improvement tolerance used as the stopping citerion. Figure \ref{idealRDcurves} shows a comparison over the RD curves used for a single clip as an example. The gains are larger at lower bitrates as expected. In this example, the BR-Rate gain was 2.5\% and Brent's method took the lowest number of iterations : 8.

\begin{figure}
    \centering

    \includegraphics[width=0.85\columnwidth]{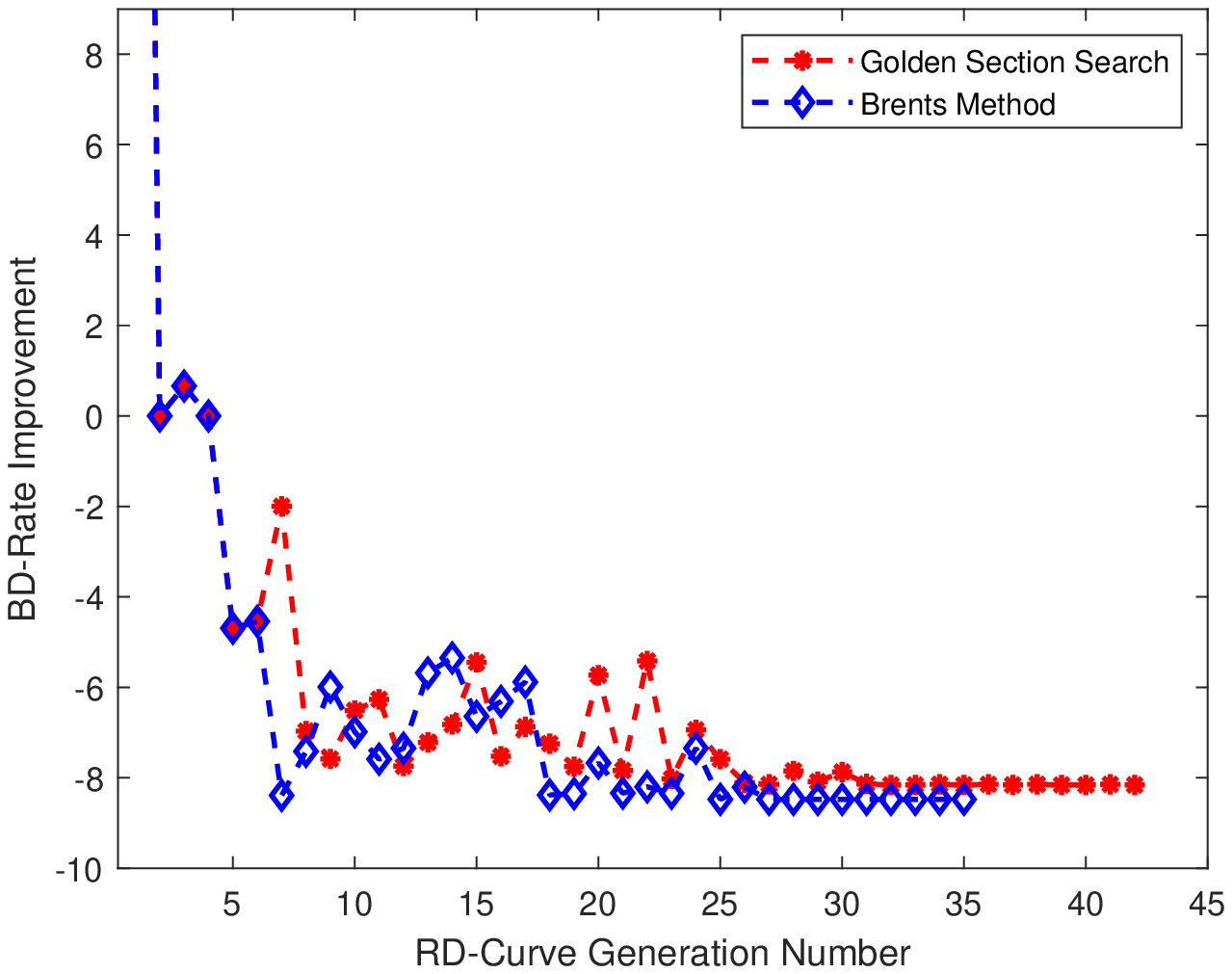}
   \includegraphics[width=0.85\columnwidth]{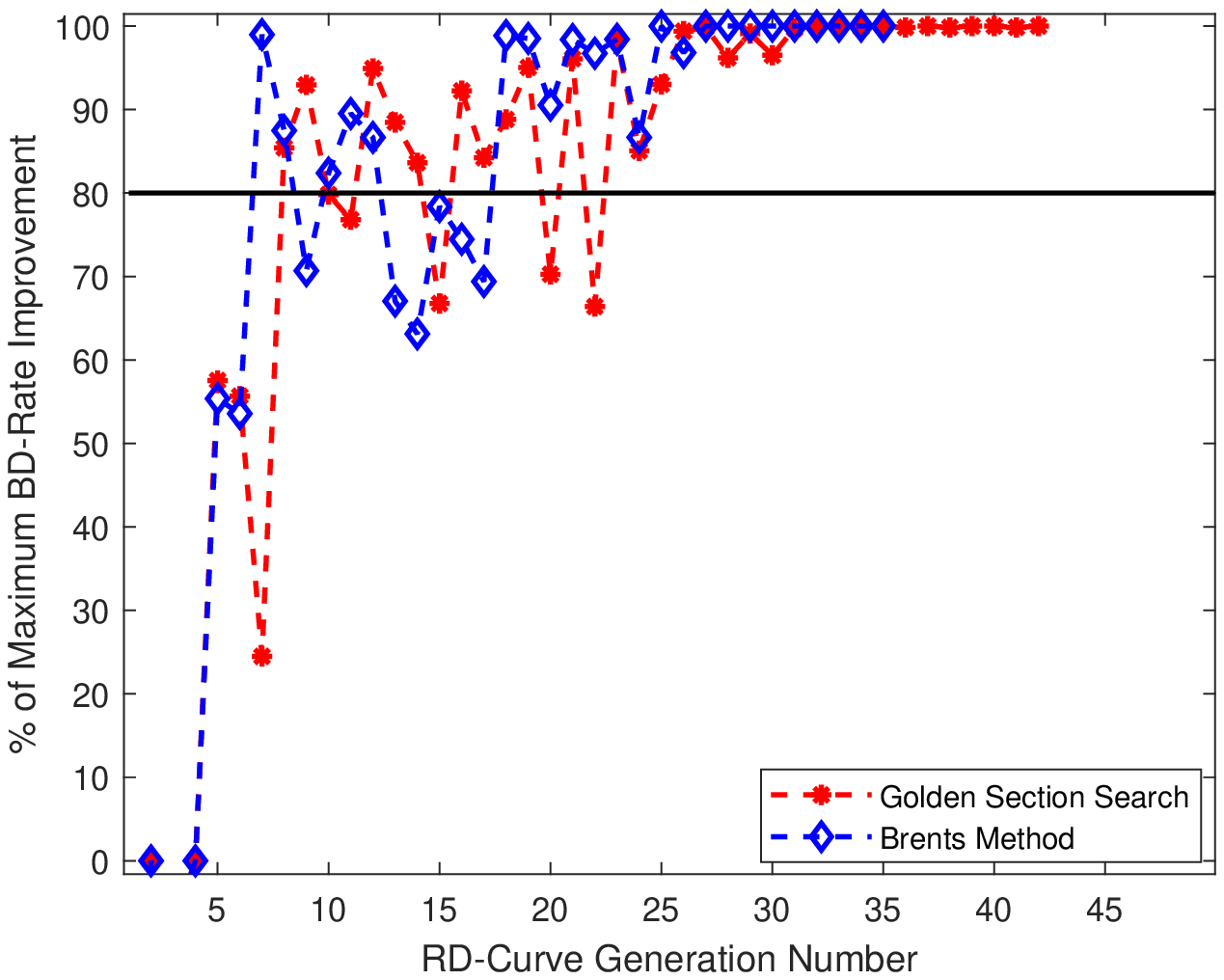}
    \caption{These two plots illustrate the convergence of the optimisation methods used for a particular clip (Sports\_720P-0b9e). Top : BD-Rate. Bottom : Normalised BD-Rate improvement. The normalised plot shows the BD-Rate improvement as a \% of the maximum BD-Rate possible for this clip. Hence it shows that 80\% of the possible BD-Rate improvement is attained within the first 15 iterations.}
    \label{minimizationExample}
\end{figure}

Videos which were encoded without the stopping criterion on the direct optimiser had an additional improvement in BD-Rate. While a further increase in compression performance is welcome, it comes with a heavy computational cost. Figure \ref{minimizationExample} shows an example of convergence in the direct optimisers in this case without the stringent stopping criterion. As can be seen the BD-Rate gains over 8\% after 35 iterations. But the lower plot shows the amount of gains achieved after each iteration as a fraction of the final converged rate. Roughly 80\% of the potential maximum BD-Rate improvement can be achieved within the first fifteen(15) iterations, therefore the stopping criterion are not that detrimental to performance especially in the light of computational load. 

\begin{figure}
    \centering

   \includegraphics[width=\columnwidth]{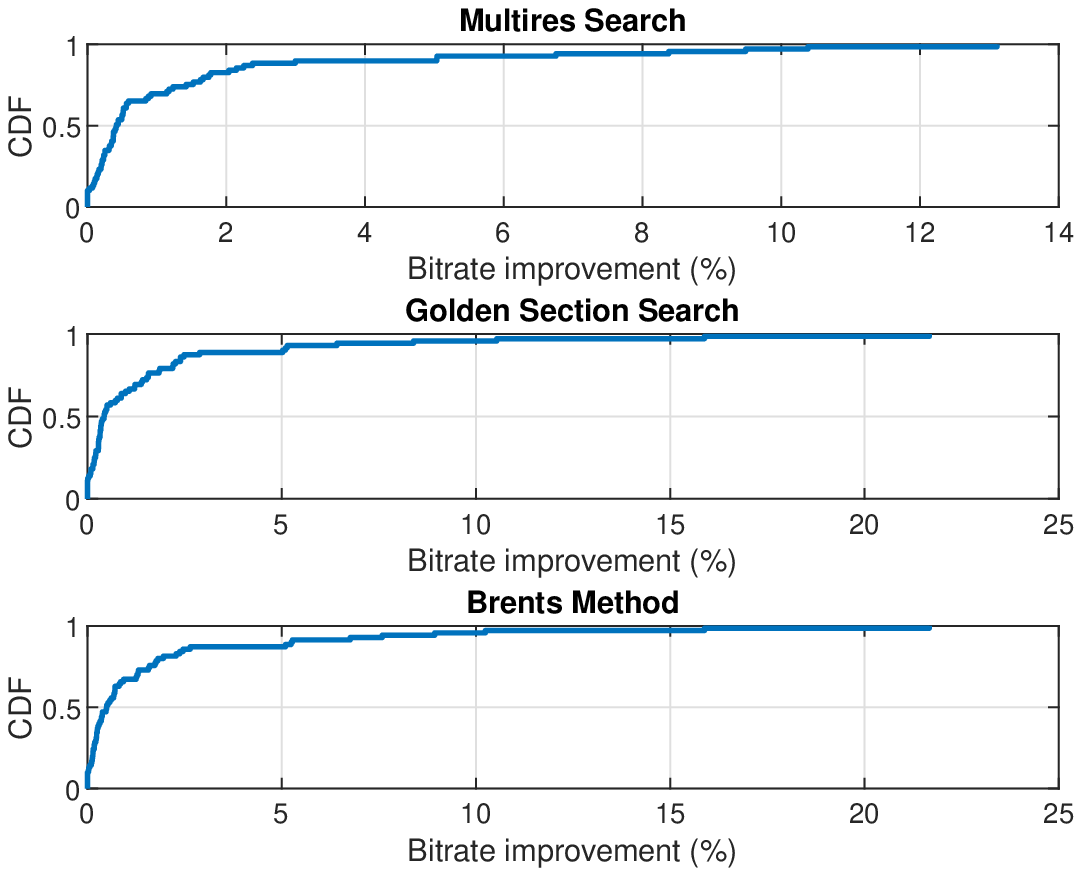}
   \includegraphics[width=\columnwidth]{avgBDRate2.eps}
   
    \caption{CDF vs bitrate improvement (\%) at 40dB. Top to bottom: Multi-res, Golden Section Search, Brents Method. At a target quality of 40 dB, 80\% of the clips had gains up to
5\% improvement for Golden Section and Brents Method }
    \label{bitrateHist}
\end{figure}

{}
In practice, we are interested in the rate gains {\em at a particular} quality. For example a 40dB quality target is a good rule of thumb for a streaming media service. BD-Rate measures instead the average gain across the RD performance curve. We can show our results in this light. Figure \ref{bitrateHist} is a cumulative histogram of the percent bitrate improvement at 40 dB for the 77 clips encoded in this paper. For 80\% of the clips we can have up to 5\% improvement at this quality. This is a useful metric and further validates the proposed methods.

{}

Table \ref{ZhangResults} compares the results from this work with the results from Zhang and Bull \cite{zhang_bull} applied to the dataset in this paper. The comparison is not strictly fair because we do not adapt the $k$ every GOP and instead use a single GOP for all the clips. However the results further reiterate the difference in the shapes of the curves {\em per clip}. As can be seen our proposed algorithm estimates values of $k$ that are quite different, and subsequently the BD-Rate improvement is much better than that using $k$ as equation \ref{zhangbull}. 

\begin{table}[!h]
\caption{Table \ref{ZhangResults}: Comparison of the BD-Rate improvement using an estimated $k$ from \cite{zhang_bull} vs $k$ determined by our direct search (Brent's method). As can be seen our estimated values are quite different and lead to much improved performance. However our adaptation scheme is different from that in Zhang and Bull. Negative BD-R implies worse performance than the unmodified codec}
\label{ZhangResults}
\begin{center}

\begin{tabular}{|l|l|l|l|l|}
\hline
\textbf{Clip}             & \textbf{$\textrm{k}_\textrm{ZB}$} & \textbf{BD-R} & \textbf{$\textrm{k}_\textrm{Ours}$} & \textbf{BD-R} \\ \hline
Animation4268      &        3.173                               &-5.81                         & 0.777                    & 0.591                     \\ \hline
CoverSong7360      &        3.444                               &-4.23                         & 1.071                    & 0.178                     \\ \hline
Gaming6403         &        3.132                               &-6.18                         & 1.438                    & 0.683                     \\ \hline
HowTo21c6          &        2.938                               &-3.15                         & 0.438                    & 5.239                     \\ \hline
LiveMusic6452      &        3.632                               &-0.85                         & 1.145                    & 0.234                     \\ \hline
LyricVideo739a     &        2.491                               &0.89                         & 1.699                    & 1.412                     \\ \hline
MusicVideo3285     &        4.298                               &-10.21                         & 1.120                    & 0.068                     \\ \hline
NewsClip7745       &        2.571                               &-2.67                         & 0.547                    & 2.482                     \\ \hline
Sports0b9e         &        3.108                               &1.89                         & 1.784                    & 8.391                     \\ \hline
TVClip02b8 &        2.112                               &-3.12                         & 0.587                    & 2.887                     \\ \hline
Vlog11c5           &        3.579                               &-2.12                         & 1.096                    & 0.166                     \\ \hline
\end{tabular}
\end{center}
\end{table}

\section{Conclusion}

This paper has presented a direct optimisation technique for Lagrange multiplier estimation in a codec. The approach in some sense represents an upper bound for what can be achieved by adjusting the Lagrangian multiplier. We report gains up to $21$\% on a particular clip and $0.5-5$\% per class without affecting the quality. At a target quality of 40 dB, 80\% of the clips had gains up to 5\%. Furthermore, the work here has used a much more realistic database of content provided by YouTube and for a larger number of clips. The results also show that the relationship between $k$ and BD-Rate varies substantially between clips, and that the region near the minimum is not a simple surface. These observations point to further work in {\em per clip} optimisation using more sophisticated machine learning techniques that exploit deeper features in the content.






\small

\bibliographystyle{ieeetr}

\bibliography{main.bib}


\begin{biography}


Daniel J. Ringis is a PhD researcher at Trinity College Dublin. He is from Trinidad and Tobago, and previously studied Electrical and Computer Engineering at the University of the West Indies, St. Augustine. His current research is focused on video compression. 

Fran\c{c}ois Piti\'e is an Ussher Assistant Professor in Media Signal Processing in the School of Engineering and part of the ADAPT Research Centre. His research is concerned with media signal processing applications related to film postproduction and film restoration. The aim of his research is to develop tools that can enhance the cinematic experience of immersion.

Anil Kokaram received the Ph.D. degree from Cambridge University, U.K., in 1993. Then, he was a Research Fellow with the Cambridge University Engineering Department and Churchill College Cambridge. In 1998, he established the Sigmedia Group (www.sigmedia.tv) at Trinity College Dublin. In 2011 a company he founded, Green Parrot Pictures, operating in the consumer image processing space, was acquired by Google. He is currently working as a Tech. Lead in the Chrome Media Group at Google, Mountain View CA. He continues to hold his Professorship with the Department of Electronic Engineering, Trinity College Dublin and is a Fellow of the College. He is the author of over 100 refereed papers in various conferences and journals over the years. His research interests include the general area of digital video processing and algorithms for stereoscopic quality improvement and video processing in the life sciences. He is best known for his work in automated motion picture restoration. 
\end{biography}

\end{document}